\documentclass[12pt]{article}

\usepackage{amssymb}
\usepackage{amsmath}
\usepackage{graphicx}

\usepackage{multirow}

\def\be{\begin{eqnarray}}
\def\ee{\end{eqnarray}}
\def\eps{\varepsilon}

\def\calB{{\cal{B}}}

\begin{document}

\title{Two-parametric fractional statistics models for anyons}
\author{Andrij Rovenchak\\
Department for Theoretical Physics,\\
Ivan Franko National University of Lviv\\
12, Drahomanov St., Lviv, UA-79005, Ukraine\\
phone: +380 32 2614443; 
e-mail: andrij.rovenchak@gmail.com
}

\maketitle

\abstract{
In the paper, two-parametric models of fractional statistics are proposed in order to determine the functional form of the distribution function of free anyons. From the expressions of the second and third virial coefficients, an approximate correspondence is shown to hold for three models, namely, the nonextensive Polychronakos statistics and both the incomplete and the nonextensive modifications of the Haldane--Wu statistics. The difference occurs only in the fourth virial coefficient leading to a small correction in the equation of state. For the two generalizations of the Haldane--Wu statistics, the solutions for the statistics parameters $g,q$ exist in the whole domain of the anyonic parameter $\alpha\in[0;1]$, unlike the nonextensive Polychronakos statistics. It is suggested that the search for the expression of the anyonic distribution function should be made within some modifications of the Haldane--Wu statistics.
}

\section{Introduction}
The possibility for particles in two dimensions to obey statistics different from both Bose and Fermi statistics was first demonstrated by Leinaas and Myrheim in 1977 \cite{Leinaas&Myrheim:1977}. Namely, they showed that, due to topological peculiarities of the two-dimensional (2D) motion, an exchange of two particles yields the wave-function phase change not necessarily equal to $\pm1$:
$\psi(x_2,x_1)=e^{i\pi\alpha}\psi(x_1,x_2)$, where $\alpha$ can be any number -- hence `anyons', the term coined by Wilczek in 1982 \cite{Wilczek:1982}.

Following the discovery of the fractional quantum Hall effect \cite{Tsui_etal:1982}, several theories were proposed for this phenomenon. The models of excitations with a fractional charge were applied in particular by Laughlin \cite{Laughlin:1983}, Halperin \cite{Halperin:1984}, Arovas \textit{et al.}\ \cite{Arovas_etal:1994}. Anyons were considered as possible candidates for the respective quasiparticles \cite{Halperin:1984,Arovas_etal:1994}.

The generalized Pauli principle suggested by Haldane  \cite{Haldane:1991} and sub\-sequent derivation of the distribution function in such a new statistics by Wu \cite{Wu:1994} initiated studies of what became known as the fractional exclusion statistics (FES). It appeared to describe the anyons on the lowest Landau level \cite{Dasnieres&Ouvry:1994}, though no complete correspondence between FES and the anyonic statistics can be established.

In recent years, some studies appeared with a prospect of future experimental observations of anyons. In particular, an interference pattern relevant to Laughlin quasiparticles was obtained by Camino \textit{et al.}\ \cite{Camino_etal:2005}. Experimental setup to observe anyons in a system consisting of a superconducting film on a semiconductor heterotransition was suggested by Weeks \textit{et al.}\ \cite{Weeks_etal:2007} and another one involving one-dimensional optical lattices was proposed by Keilmann \textit{et al.}\ \cite{Keilmann_etal:2011}. Recent developments in the field of anyonic and fractional statistics also include studies of the so-called non-Abelian anyons \cite{Polychronakos:2000,You_etal:2013,Mancarella_etal:2013} being the analogs of Green's parastatistics \cite{Green:1953} and applications of the exclusion statistics to various systems, cf.\ \cite{Lundholm&Solovej:2013,Mandal:2013}.

Description of anyons within the statistical mechanics cannot be considered complete so far. Even a system of non-interacting anyons is challenging due to the presence of statistical interactions \cite{Engquist:2009}. The issue of finding an expression for the distribution function in the anyonic statistics is an open one: it is possible to establish some approximate correspondences with several known fractional statistics types using the second virial coefficient but all of them fail to catch already the third virial coefficient correctly \cite[Chap.~5]{Khare:2005}. A two-parametric statistics can be an option providing a more accurate statistical mechanical model. It must be noted that two-parametric generalizations of statistics are somewhat underrepresented in the literature. In most cases, they are based on the so-called $q$-deformations of the commutation relations between the creation and annihilation operators, cf.\  \cite{Dalton&Inomata:1995,Algin_etal:2002,Algin:2010,Gavrilik&Mishchenko:2013}. Some generalizations having common features with those considered in the present work were briefly discussed by Kaniadakis \cite{Kaniadakis:2001}, albeit from different grounds.

A two-fold modification of statistics applied in this paper is obtained as follows. First, the counting of microstates is approached either in a manner suggested by Polychronakos \cite{Polychronakos:1996} or within the fractional exclusion statistics by Haldane and Wu \cite{Haldane:1991,Wu:1994}. This leads to a functional form of the distribution function different from that in the Bose/Fermi statistics and introduces one parameter into the model. Next, a deformation of the exponential in the Gibbs factor is applied giving a second statistics parameter. For this purpose, the Tsallis $q$-exponential can be used. Indeed, the nonextensive statistical mechanics suggested by Tsallis \cite{Tsallis:1988} is applicable in particular to systems with long-range interactions \cite{Abe&Okamoto:2001} -- and there exists such a statistical inter\-action in a system of anyons \cite{Wilczek:1982}. Note, however, that the deformation of exponentials is made here phenomenologically and the so-called incomplete statistics \cite{Wang:2003,Kaupp_etal:2008} is also considered.

\section{Virial and cluster expansions}
Let us first briefly recall the connection between the virial and cluster expansions in statistical mechanics to be applied here for 2D systems.

With $P$ being the pressure and $T$ denoting the temperature (for simplicity, Boltzmann's constant is put equal to unity), the equation of state for a two-dimensional system in the low-density, high-temperature limit can be written as the following virial expansion: 
\be
\frac{P}{T} = \rho_2 \left[1+ b_2\,\rho_2\lambda^2 
+ b_3\, (\rho_2\lambda^2)^2+\ldots\right],
\ee
where the two-dimensional density $\rho_2 = N/A$ with $N$ being the number of particles and $A$ standing for the area; $b_2, b_3, \ldots$ are the dimensionless virial coefficients and $\lambda$ is the thermal de~Broglie wavelength of a particle with mass $m$
\be
\lambda = \sqrt{\frac{2\pi\hbar^2}{mT}}.
\ee

On the other hand, the grand-canonical partition function $\Xi$ can be written as a series over the fugacity $z=e^{\mu/T}$, where $\mu$ is the chemical potential, as follows:
\be
\frac{1}{A}\ln\Xi = \sum_{\ell=1}^\infty \calB_\ell z^\ell.
\ee
The coefficients $\calB_\ell$ are known as cluster integrals. They can be used to calculate virial coefficients in view of the thermodynamic relations linking the pressure and density with the grand-canonical partition function, namely:
\be
\frac{P}{T} = \frac{1}{A}\ln\Xi
\ee
and
\be
\rho_2 = \frac{N}{A} = z\frac{\partial}{\partial z}
\left(\frac{1}{A}\ln\Xi\right)_{A,T}.
\ee
After some transformations, the following relations for the virial coefficients are obtained \cite{Borges_etal:1999}:
\be
\label{b2-def}
&& b_2 \lambda^2 = -\frac{\calB_2}{\calB_1^2},\\
\label{b3-def}
&& b_3 \lambda^4 = -2\frac{\calB_3}{\calB_1^3} + 4\frac{\calB_2^2}{\calB_1^4},\\
\label{b4-def}
&& b_4 \lambda^6 = 
-3\frac{\calB_4}{\calB_1^4} + 18\frac{\calB_2\calB_3}{\calB_1^5}
-20\frac{\calB_2^3}{\calB_1^6},\\
&& \ldots\ . \nonumber
\ee

Aiming to find an expression for the distribution function (oc\-cu\-pa\-tion number) of anyons $n_j$, such that
\be\label{N=sumj}
N = \sum_j g_j n_j,
\ee
where the sum runs over all the energy levels with degeneracies $g_j$, the following equation is easily derived:
\be\label{NA=sumj}
\frac{N}{A} = \frac{1}{A} \sum_j g_j n_j = 
\sum_{\ell=1}^\infty \ell\calB_\ell z^\ell.
\ee
As the functional dependence of $n_j$ on the level energy $\eps_j$ is not known, in the next Section several modifications of fractional statistics are suggested and checked for the suitability to model anyons basing on the expressions for the second and third virial coefficients.

\section{Modifications of statistics}
Further in this work, the modifications of fractional statistics are based on the following two general expressions. The first one,
\be\label{nj-P}
n_j^{\rm P} = \frac{1}{z^{-1}X(\eps_j) + Y},
\ee
generalizes standard Bose (Fermi) statistics determined by $X(\eps_j)=e^{\eps_j/T}$ and $Y=-1$ ($Y=+1$) and the fractional Polychronakos statistics with $Y=-\gamma={\rm const}\neq\pm1$.

The second modification is represented by the Haldane--Wu statistics with
\be\label{nj-HW}
n_j^{\rm HW} = \frac{1}{w[z^{-1}X(\eps_j)] +g},
\ee
where the function $w(\xi)$ is the solution to such a transcendental equa\-tion:
\be
w^g(1+w)^{1-g} = \xi,
\ee
which is obtained using an expression for the number of microstates of a quantum many-body system being a simple interpolation between bosons and fermions \cite{Wu:1994}.

Additionally to the $\gamma$ parameter in the Polychronakos statistics (\ref{nj-P}) and the $g$ parameter in the Haldane--Wu statistics (\ref{nj-HW}), a second parameter can be introduced into the model by a deformation of the exponential defining the $X$ dependences in Eqs~(\ref{nj-P})--(\ref{nj-HW}). Two such deformations are considered in this work, namely, the Tsallis $q$-exponential $e_q^x$ appearing instead of the ordinary $e^x$ in the nonextensive statistics and the change of $e^x$ to $e^{qx}$ known in the incomplete statistics \cite{Wang:2003,Kaupp_etal:2008}.

While the deformation of exponentials in the Gibbs factor $X=e^{\eps_j/T}$ using the $q$ parameter is made phenomenologically here, a physical motivation is provided for such a procedure as follows. A long-range statistical interaction in the system of anyons \cite{Wilczek:1982} can contribute to the nonextensivity, which is taken into consideration within the Tsallis approach \cite{Tsallis:1988,Tsallis:1994}. On the other hand, some correspondence between the incomplete and the Tsallis statistics can be established \cite{Nivanen_etal:2005}.

The Tsallis $q$-exponential is given by \cite{Tsallis:1994}
\be
e_q^x = [1 + (1 - q)x]^{1/(1-q)} &&\quad\textrm{for}\ 1 + (1 - q)x \geq 0\\
\nonumber
\textrm{and} && e_q^x = 0\ \textrm{otherwise}.
\ee
In the incomplete statistics, the normalization of probabilities $p_i$ is given by \cite{Wang:2003,Kaupp_etal:2008}
\be
\sum_i p_i^q = 1
\qquad\textrm{instead of}\qquad
\sum_i p_i = 1.
\ee
As it is easily seen, both cases reduce to standard exponentials when $q=1$.

The small-$z$ expansions of Eqs.~(\ref{nj-P}) and (\ref{nj-HW}) are given by:
\be
n_j^{\rm P} &=& \sum_{\ell=1}^\infty 
(-1)^{\ell-1}\frac{Y^{\ell-1}}{X^\ell}\,z^\ell
=\frac{1}{X}z - \frac{Y}{X^2}z^2 + \frac{Y^2}{X^3}z^3\mp\ldots,
\label{nP-def}\\
n_j^{\rm HW} &=& \sum_{m=0}^\infty 
(-1)^m\frac{\Gamma[g(m+1)]}{m!\Gamma[g(m+1)-m]} 
\frac{z^{m+1}}{X^{m+1}}\nonumber\\
&=&
\frac{1}{X}z - \frac{(2g-1)}{X^2}z^2 + \frac{(3g-2)(3g-1)}{2!\,X^3}z^3\mp\ldots,
\label{nHW-def}
\ee
cf. \cite{Delves_etal:1997}.

For simplicity, the summation over the levels $\sum_j$ in Eqs.~(\ref{N=sumj})--(\ref{NA=sumj}) can be substituted with the integration over energies upon introducing the density of states function $G(\eps)$, which for a 2D ideal gas of particles having the mass $m$ is $G(\eps) = mA/2\pi\hbar^2={\rm const}$:
\be
\sum_j \ldots = \int_0^\infty d\eps\, G(\eps)\ldots.
\ee  
Using different representations for the $X$ function, cluster integrals $\calB_\ell$ are then straightforward to calculate from expansion (\ref{nP-def})--(\ref{nHW-def}) giving in turn virial coefficients by virtue of Eqs.~(\ref{b2-def})--(\ref{b4-def}).

\section{Results}
\textit{Incomplete Polychronakos statistics (IPS).} For this type of statistics, 
$X(\eps) = e^{q\eps/T}$, $Y = -\gamma$, yielding cluster integrals
\be
\calB_1 \lambda^2 = \frac{1}{q},\qquad
\calB_2 \lambda^2 = \frac{\gamma}{4q},\qquad
\calB_3 \lambda^2 = \frac{\gamma^2}{9q},\qquad\ldots
\ee
and the second and third virial coefficients are
\be
b_2^{\rm IPS} = -\frac{\gamma q}{4},\qquad
b_3^{\rm IPS} = \frac{\gamma^2 q^2}{36}.
\ee

\textit{Nonextensive Polychronakos statistics (NEPS).} For this type of statistics, 
$X(\eps) = e_q^{\eps/T}$, $Y = -\gamma$, yielding cluster integrals
\be
\calB_1 \lambda^2 = \frac{1}{q},\qquad
\calB_2 \lambda^2 = \frac{\gamma}{2(1+q)},\qquad
\calB_3 \lambda^2 = \frac{\gamma^2}{3(2+q)},\qquad\ldots
\ee
and the second and third virial coefficients are
\be\label{b2,3-NEPS}
b_2^{\rm NEPS} = -\frac{\gamma q^2}{2(1+q)},\qquad
b_3^{\rm NEPS} = 
\gamma^2q^4\left[\frac{1}{(1+q)^2}-\frac{1}{3q(2+q)}\right].
\ee
Note that this statistics in the limits of weak nonextensivity $q\to1$ and 
$\gamma\to1$ can be applied to model a finite weakly-interacting Bose-system \cite{Rovenchak:2014b}.

\textit{Incomplete Haldane--Wu statistics (IHWS).} For this type of statistics, 
$X(\eps) = e^{q\eps/T}$, yielding cluster integrals
\be
\calB_1 \lambda^2 = \frac{1}{q},\quad
\calB_2 \lambda^2 = -\frac{(2g-1)}{2! \cdot 2q},\quad
\calB_3 \lambda^2 = \frac{(3g-2)(3g-1)}{3! \cdot 3q},\ \ldots
\ee
and the second and third virial coefficients are
\be\label{b2,3-IHWS}
b_2^{\rm IHWS} = \frac{(2g-1)q}{4},\qquad
b_3^{\rm IHWS} = \frac{q^2}{36}.
\ee

\textit{Nonextensive Haldane--Wu statistics (NEHWS).} For this type of statistics, 
$X(\eps) = e_q^{\eps/T}$, yielding cluster integrals
\be
\calB_1 \lambda^2 = \frac{1}{q},\quad
\calB_2 \lambda^2 = -\frac{(2g-1)}{2(1+q)},\quad
\calB_3 \lambda^2 = \frac{(3g-2)(3g-1)}{6(2+q)},\ \ldots
\ee
and the second and third virial coefficients are
\be\label{b2,3-NEHWS}
&&b_2^{\rm NEHWS} = \frac{2g-1}{2}\frac{q^2}{1+q},\nonumber\\
\\
&&b_3^{\rm NEHWS} = 
q^4\left[\frac{(2g-1)^2}{(1+q)^2}-\frac{(3g-2)(3g-1)}{3q(2+q)}\right].
\nonumber
\ee

To find the fractional statistics with the distribution function best fitting to the anyonic statistics, the above expressions in each statistics for $b_2$ and $b_3$ containing two parameters are equated to the respective virial coefficients of anyons providing sets of two equations. The solutions of these sets allow judging which of the analyzed models is appropriate.
 
The second and third virial coefficients of an ideal anyon gas are given by \cite{Mashkevich_etal:1996}:
\be\label{b2,3-anyon}
b_2^{\rm anyon} = -\frac{1}{4}(1-4\alpha+2\alpha^2),\quad
b_3^{\rm anyon} = \frac{1}{36}+\frac{\sin^2\pi\alpha}{12\pi^2}+
c_3\sin^4\pi\alpha,\nonumber\\
\\
c_3 = -(1.652\pm0.012) \cdot 10^{-5},\nonumber
\ee
where the statistics parameter $\alpha\in[0;1]$ interpolates between the Bose statistics ($\alpha=0$) and Fermi statistics ($\alpha=1$). The above expression for the second virial coefficient is exact while in $b_3^{\rm anyon}$ higher harmonics ($\sin^6\pi\alpha$ etc.) are neglected.

For the simplest of the proposed models, the incomplete Poly\-chro\-nakos statistics, the solutions of $\left\{b_2^{\rm IPS} = b_2^{\rm anyon},\ b_3^{\rm IPS} = b_3^{\rm anyon}\right\}$ exist only in the trivial cases $\alpha=0,1$ corresponding the Bose and Fermi statistics, respectively.

A wider domain of the $\alpha$ values, but still not the entire $[0;1]$ segment, can be reproduced by the nonextensive Polychronakos statistics. From (\ref{b2,3-NEPS}) one can obtain the condition defining the unaccessible values of the anyonic statistics parameter $\alpha\in(\alpha_1;\alpha_2)$ as follows:
\be
\frac{b_3^{\rm anyon}(\alpha_{1,2})}{\left[b_2^{\rm anyon}(\alpha_{1,2})\right]^2} = \frac{4}{3},
\ee
which is derived as the $q\to\infty$ limit, giving 
\be
\alpha_1 = 0.109\ldots; \qquad \alpha_2 = 0.584\ldots\;.
\ee
The nonextensive Polychronakos statistics can thus be used to model anyons on the bosonic side, $0\leq\alpha <\alpha_1$, and on the fermionic side, $\alpha_2< \alpha \leq 1$. The correspondence between the $\gamma,q$ parameters and $\alpha$ values is presented in Table~\ref{tab:params}.

\begin{figure}[h!]
\centerline{\includegraphics[scale=0.720]{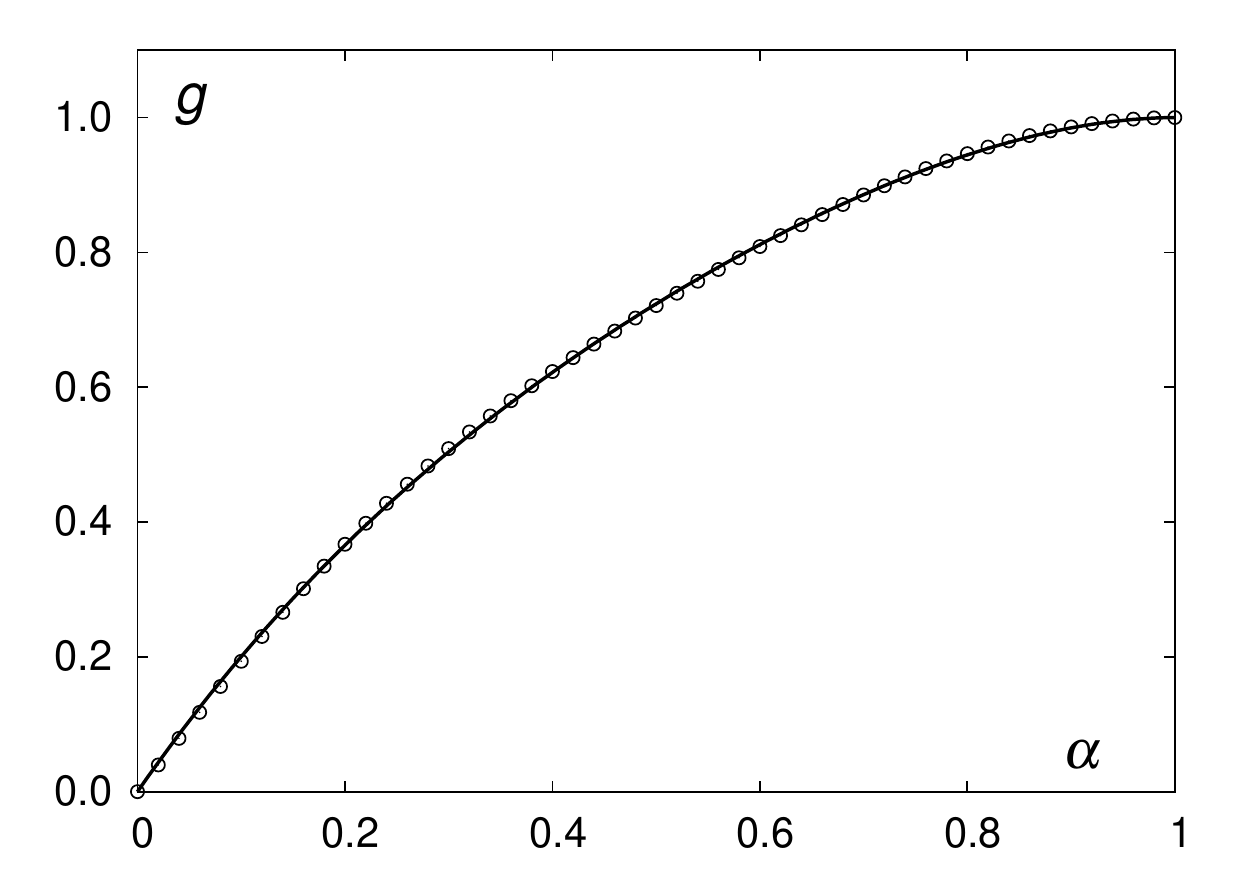}}
\caption{The $g$ parameter of the nonextensive Haldane--Wu statistics (circles) shown with the fitting function (solid line). The following simple polynomial was used: 
$f(\alpha) =  a_1 \alpha + a_2 \alpha^2 + a_3 \alpha^3 + (1-a_1-a_2-a_3)\alpha^4$ with
$a_1 = 2.21 \pm 0.01$, 
$a_2 = -2.26 \pm 0.05$,
$a_3 = 1.88 \pm 0.09$.
}
\label{fig:g-alpha}
\end{figure}

\begin{figure}[h!]
\centerline{\includegraphics[scale=0.720]{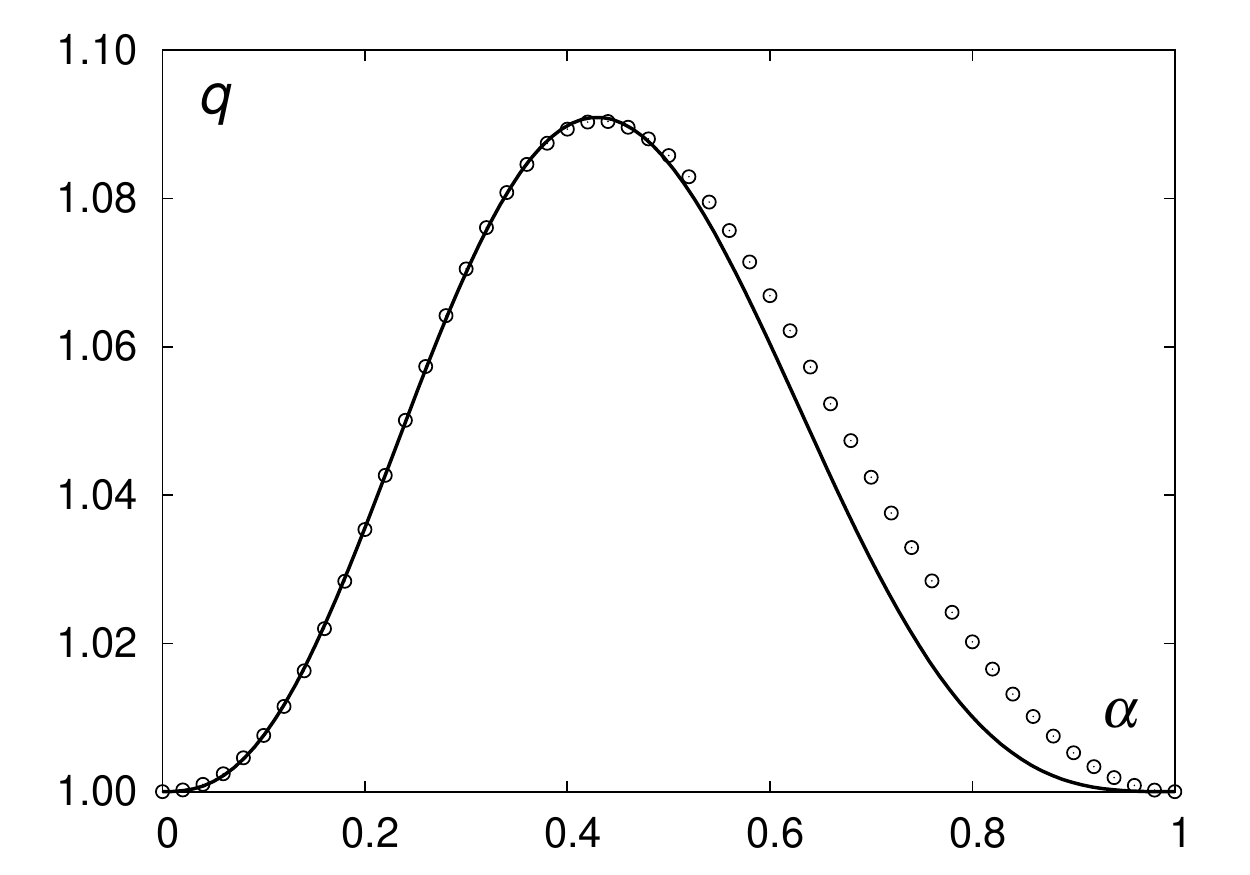}}
\caption{The $q$ parameter of the nonextensive Haldane--Wu statistics (circles) shown with the fitting function (solid line). The following function was fitted to the data for $\alpha\in[0;0.5]$: 
$f(\alpha) = 1+ b\sin^k (\pi \alpha^s)$ with
$b = 0.09096 \pm 0.00009$,
$s = 0.820 \pm 0.001$,
$k = 3.18 \pm 0.02$.
}
\label{fig:q-alpha}
\end{figure}

The proposed modifications of the Haldane--Wu statistics appear to suit the anyonic statistics better than that of the Polychronakos statistics discussed above. In its incomplete version, see (\ref{b2,3-IHWS}), very simple relations of the statistics parameters $g$ and $q$ with $\alpha$ are achieved:
\be
q &=& \left(1+\frac{3}{\pi^2}\sin^2\pi\alpha+36c_3\sin^4\pi\alpha\right)^{1/2},
\nonumber\\
\\
1-2g &=& \frac{1}{q}(1-4\alpha+2\alpha^2).
\nonumber
\ee
 In the nonextensive modification, see (\ref{b2,3-NEHWS}), the analytical expressions are rather cumbersome but the links between $g,q$ and $\alpha$ are easily calculated numerically, see Figs.~\ref{fig:g-alpha},\ref{fig:q-alpha} for details. The correspondence between the $g,q$ parameters and $\alpha$ values in both IHWS and NEHWS is shown in Table~\ref{tab:params}.

\bigskip

\begin{table}[h]
\caption{Dependence of the parameters of the nonextensive Polychronakos statistics, incomplete and nonextensive Haldane--Wu statistics on the anyonic parameter $\alpha$}\label{tab:params}
\begin{center}
\begin{tabular}{ccccccc}
\hline\hline
\multirow{2}{*}{$\alpha$}  &\multicolumn{2}{|c|}{NEPS} &
\multicolumn{2}{c|}{IHWS} &
\multicolumn{2}{c}{NEHWS}\\
&\multicolumn{1}{|c}{$\gamma$}	& \multicolumn{1}{c|}{$q$}	
&	$g$	&	\multicolumn{1}{c|}{$q$}	 &	$g$	&	$q$	\\
\hline
\vphantom{$\int\limits^.$}%
0.00	&	1.00000	&	1.00000	&	0.00000	&	1.00000	&	0.00000	&	1.00000	\\
0.01	&	0.91469	&	1.03299	&	0.01997	&	1.00015	&	0.01994	&	1.00006	\\
0.05	&	0.56672	&	1.26962	&	0.09899	&	1.00371	&	0.09847	&	1.00161	\\
0.10	&	0.11697	&	3.42415	&	0.19440	&	1.01441	&	0.19349	&	1.00758	\\
0.20	&	---	&	---	&	0.36681	&	1.05116	&	0.36709	&	1.03536	\\
0.30	&	---	&	---	&	0.50913	&	1.09485	&	0.50903	&	1.07051	\\
0.40	&	---	&	---	&	0.62401	&	1.12892	&	0.62325	&	1.08934	\\
0.45	&	---	&	---	&	0.67349	&	1.13840	&	0.67369	&	1.09008	\\
0.50	&	---	&	---	&	0.71898	&	1.14165	&	0.72115	&	1.08580	\\
0.55	&	---	&	---	&	0.76133	&	1.13840	&	0.76612	&	1.07765	\\
0.60	& $-0.10436$ &	4.06034	&	0.80117	&	1.12892	&	0.80869	&	1.06690	\\
0.65	& $-0.30776$ &	1.87929	&	0.83888	&	1.11397	&	0.84859	&	1.05479	\\
0.70	& $-0.48484$ &	1.43495	&	0.87448	&	1.09485	&	0.88532	&	1.04241	\\
0.80	& $-0.76642$ &	1.13091	&	0.93761	&	1.05116	&	0.94643	&	1.02020	\\
0.90	& $-0.94088$ &	1.02760	&	0.98304	&	1.01441	&	0.98618	&	1.00524	\\
0.95	& $-0.98517$ &	1.00664	&	0.99566	&	1.00371	&	0.99652	&	1.00132	\\
0.99	& $-0.99941$ &	1.00026	&	0.99983	&	1.00015	&	0.99986	&	1.00005	\\
1.00	& $-1.00000$	&	1.00000	&	1.00000	&	1.00000	&	1.00000	&	1.00000	\\
\hline\hline
\end{tabular}
\end{center}
\end{table}

The accuracy of the proposed models can be tested using the fourth virial coefficient. For anyons, it equals \cite{Kristoffersen_etal:1998}:
\be\label{b4-anyon}
&&\hspace*{-0.5cm}
b_4^{\rm anyon} = \frac{\sin^2\pi\alpha}{16\pi^2}
\left(\frac{\ln(\sqrt{3}+2)}{\sqrt{3}} + \cos\pi\alpha\right)+
(c_4+d_4\cos\pi\alpha)\sin^4\pi\alpha,
\nonumber\\
\\
&& c_4 = -0.0053\pm0.0003;\qquad d_4 = -0.0048\pm 0.0009.
\nonumber
\ee
This virial coefficient appears to have small values in the domain $-0.0006<b_4^{\rm anyon}<0.0025$. 

\begin{figure}[h]
\centerline{\includegraphics[scale=0.720]{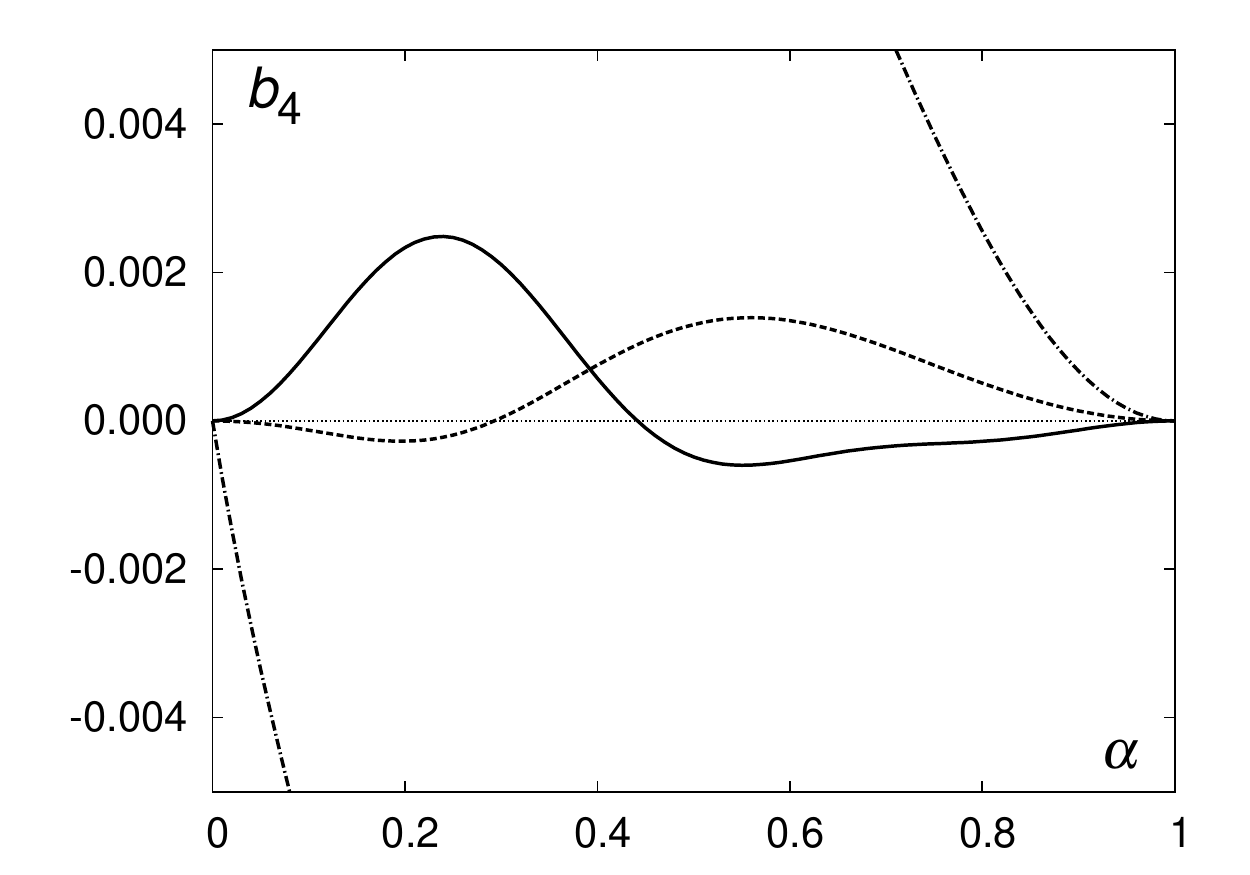}}
\caption{The fourth virial coefficient of free anyons (solid line) compared to that in the nonextensive modifications of the Polychronakos statistics (dashed-dotted line) and Haldane--Wu statistics (dashed line).}
\label{fig:b4}
\end{figure}

In the incomplete Haldane--Wu statistics, $b_4^{\rm IHWS}= 0$ for all the values of the statistics parameters. In the nonextensive modification, $-0.0003<b_4^{\rm NEHWS}<0.0014$ being thus also a small number. However, the behavior of the fourth virial coefficient between the Bose and Fermi limits is not reproduced correctly in the proposed variants of statistics, see Fig.~\ref{fig:b4}. It means that the correspondence of the discussed two-parametric fractional statistics types with the anyonic one is established only approximately, though with a better accuracy than can be achieved in a one-parametric case.

\section{Conclusions}
In summary, two-parametric models of fractional statistics were considered in the work aiming to determine the functional form of the distribution function of free anyons. Approximate correspondence was shown to hold for three models, namely, the nonextensive Polychronakos statistics and both the incomplete and the nonextensive Haldane--Wu statistics. The second and third virial coefficients were reproduced correctly and the difference occurred only in the fourth virial coefficient, which leads  to a small correction in the equation of state. For the two modifications of the Haldane--Wu statistics, the solutions for the statistics parameters $g,q$ exist in the whole domain of the anyonic parameter $\alpha\in[0;1]$, unlike the nonextensive Polychronakos statistics.

It seems thus favorable to search for the expression of the anyonic distribution function within modifications of the Haldane--Wu statistics. Possible generalizations can include in particular a different definition of the $q$-exponential \cite{Lavagno:2009} and the $\varkappa$-deformed exponential \cite{Kaniadakis:2001,Kaniadakis:2013}. The analysis of $q$-deformed Fermi and Bose statistics \cite{Algin&Senay:2012,Zeng_etal:2014} with further introduction of additional parameters is also feasible.

\section*{Acknowledgments} I am grateful to Prof.~Volodymyr Tkachuk for his feedbacks on this work. The present study was partially supported by Project $\Phi\Phi$-110$\Phi$ (registration number 0112U001275) from the Ministry of Education and Science of Ukraine.


\begin{thebibliography}{00}
\bibitem{Leinaas&Myrheim:1977}J. M. Leinaas and J. Myrheim, Nuovo Cim. \textbf{37B}, 1 (1977).
\bibitem{Wilczek:1982}F. Wilczek, Phys. Rev. Lett. \textbf{49}, 957 (1982).
\bibitem{Tsui_etal:1982}D.~C.~Tsui, H.~L.~Stormer, A.~C.~Gossard, Phys. Rev. Lett. \textbf{48}, 1559 (1982).
\bibitem{Laughlin:1983}R.~B.~Laughlin, Phys. Rev. Lett. \textbf{50}, 1395 (1983).
\bibitem{Halperin:1984}B.~I.~Halperin, Phys. Rev. Lett. \textbf{52}, 1583 (1984).
\bibitem{Arovas_etal:1994}D.~Arovas, J.~R.~Schrieffer, F.~Wilczek, Phys. Rev. Lett. \textbf{53}, 722 (1984).

\bibitem{Haldane:1991}F.~D.~M.~Haldane, Phys. Rev. Lett. \textbf{67}, 937 (1991).
\bibitem{Wu:1994}Y.-S.~Wu,  Phys. Rev. Lett. \textbf{73}, 922 (1994).
\bibitem{Dasnieres&Ouvry:1994}A.~Dasni\'eres de Veigy, S.~Ouvry, Phys. Rev. Lett. \textbf{72}, 600 (1994).

\bibitem{Camino_etal:2005}F.~E.~Camino, W.~Zhou, V.~J.~Goldman, Phys. Rev. B \textbf{72} 075342 (2005).
\bibitem{Weeks_etal:2007}C.~Weeks, G.~Rosenberg, B.~Seradjeh, M.~Franz, Nature Phys. \textbf{3} 797 (2007).
\bibitem{Keilmann_etal:2011}T.~Keilmann, S.~Lanzmich, I.~McCulloch, M.~Roncaglia, Nature Commun. \textbf{2}, 361 (2011).

\bibitem{Polychronakos:2000}A. P. Polychronakos, Phys. Rev. Lett. \textbf{84}, 1268 (2000).
\bibitem{You_etal:2013}Yi-Zhuang You, Chao-Ming Jian, and Xiao-Gang Wen, Phys. Rev. B \textbf{87}, 045106 (2013).
\bibitem{Mancarella_etal:2013}F.~Mancarella, A.~Trombettoni, and
G.~Mussardo, Nucl. Phys. B \textbf{867} [FS], 950 (2013).
\bibitem{Green:1953}H. S. Green, Phys. Rev. \textbf{90}, 270 (1953).
\bibitem{Lundholm&Solovej:2013}D.~Lundholm and J.~P.~Solovej, Phys. Rev. A \textbf{88}, 062106 (2013).
\bibitem{Mandal:2013}S.~Mandal, Pramana -- J. Phys. \textbf{81}, 503 (2013).

\bibitem{Engquist:2009}J.~Engquist, Nucl. Phys. B \textbf{816} [FS], 356 (2009).
\bibitem{Khare:2005}A. Khare, \textit{Fractional Statistics and Quantum Theory}, 2nd edition (World Scientific, 2005).

\bibitem{Dalton&Inomata:1995}S.~L.~Dalton and A. Inomata, Phys. Lett. A \textbf{199}, 315 (1995).
\bibitem{Algin_etal:2002}A.~Algin, M.~Arik, and A.~S.~Arikan, Phys. Rev. E \textbf {65}, 026140 (2002).
\bibitem{Algin:2010}A.~Algin, Commun. Nonlinear Sci. Numer. Simulat. \textbf{15}, 1372 (2010).
\bibitem{Gavrilik&Mishchenko:2013}A.~M.~Gavrilik and Yu.~A.~Mishchenko, Ukr. J. Phys. \textbf{58}, 1171 (2013).
\bibitem{Kaniadakis:2001}G.~Kaniadakis, Physica A \textbf{296}, 4005 (2001).

\bibitem{Polychronakos:1996}A. P. Polychronakos, Phys. Lett. B \textbf{365}, 202 (1996).

\bibitem{Tsallis:1988}C.~Tsallis, J. Stat. Phys. \textbf{52}, 479 (1988).
\bibitem{Abe&Okamoto:2001}\textit{Nonextensive Statistical Mechanics and Its Applications}, edited by \textit{S.~Abe and Y.~Okamoto} (Springer, 2001).
\bibitem{Wang:2003}Q.~A.~Wang, Entropy \textbf{5}, 220 (2003).
\bibitem{Kaupp_etal:2008}Y.~Kaupp \textit{et al.}, J. Low Temp. Phys. \textbf{150}, 660 (2008).

\bibitem{Borges_etal:1999}P.~F.~Borges, H.~Boschi-Filho, and C.~Farina, Mod. Phys. Lett. A \textbf{14} 1217 (1999).

\bibitem{Tsallis:1994}C.~Tsallis, Qu\'\i{}mica Nova \textbf{17}, 468 (1994).

\bibitem{Nivanen_etal:2005}L.~Nivanen, M.~Pezeril, Q.~A.~Wang, and A.~Le~M\'ehaut\'e, Chaos, Solitons and Fractals \textbf{24}, 1337 (2005).

\bibitem{Delves_etal:1997}R.~T.~Delves, G.~S.~Joyce, and I.~J.~Zucker, Proc. R. Soc. Lond. A \textbf{453}, 1177 (1997).

\bibitem{Rovenchak:2014b}A.~Rovenchak, e-print arXiv:1401.4029 (2014).

\bibitem{Mashkevich_etal:1996}S.~Mashkevich, J.~Myrheim, K.~Olaussen, Phys. Lett. B \textbf{382}, 124 (1996).
\bibitem{Kristoffersen_etal:1998}A.~Kristoffersen, S.~Mashkevich, J.~Myrheim, and K.~Olaussen, Int. J. Mod. Phys. A \textbf{13}, 3723 (1998).

\bibitem{Lavagno:2009}A.~Lavagno, Rep. Math. Phys. \textbf{64}, 79 (2009).

\bibitem{Kaniadakis:2013}G.~Kaniadakis, Entropy \textbf{15}, 3983 (2013).
\bibitem{Algin&Senay:2012}A.~Algin, M.~Senay, Phys. Rev. E \textbf{85}, 041123 (2012).
\bibitem{Zeng_etal:2014}Qi-Jun Zeng, Yong-Song Luo, Yuan-Guo Xu, Hao Luo, Physica A \textbf{398}, 116 (2014).
\end{thebibliography}
\end{document}